\documentstyle[11pt]{article}
 \newcommand{\beq}{\begin{equation}}                       
 \newcommand{\eeq}{\end{equation}}                         
 \newcounter{nt}[section]                                  
 \newcounter{nl}[section]                                  
 \date{}                                                   

\begin{document}

\begin{center}
{\LARGE \bf Wave Hierarchies in Viscoelasticity}
\end{center}
\vspace{4mm}

\begin{center}
 { M. DE ANGELIS, P. MASSAROTTI  {\small AND} P. RENNO}
\end{center}

\begin {center}
 {\small
  Facolt\`a di Ingegneria, Dip. di Mat. e Appl.,
 via Claudio 21, 80125, Napoli, Italy.}
\end {center}
\begin{center}
\vspace{-3mm}
 $<$ {\small modeange}$>$ $<$ {\small massarotti}$>$  $<$ {\small renno}$>${\small @unina.it}
\end{center}

\vspace{5mm}

{\footnotesize

\noindent{\bf Abstract-} An evolution operator
  $ {\cal L}_n $ with $n$ arbitrary, typical of several models, is analyzed.
  When $n=1$ the operator characterizes the Standard Linear Solid of
  viscoelasticity, whose properties are already extablished in
  previous papers. The fundamental solution  ${\cal E}_n$
of ${\cal L}_n$ is explictly obtained and it's estimated in terms of the
fundamental solution  ${\cal E}_1 $ of  ${\cal L}_1 $. So, whatever n may
be, asymptotic properties and maximum theorems are achieved. These
results are applied to {\em the Rouse model}  and {\em reptation
model},
 which describe different aspects of polymer chains.
}

  \vspace{7mm}
{\footnotesize
\noindent{\bf Keywords-} Viscoelasticity, Wave hierarchies, Partial
differential equation, Maximum principles.
}

\vspace{5mm}

\begin{center}
 \section{
{\bf \hspace*{-8mm}. \hspace*{-2mm}INTRODUCTION}
}
\end{center}

 \setcounter{equation}{0}

 \hspace{5.1mm}

Let ${\cal L}_n $ be the $(2+n)$ order operator:

\vspace{3mm}
 \beq                                   \label{11}
 {\cal L}_ n =   \partial_{t}^{(n)} (
\partial_{tt}-c_n^2 \partial_{xx} ) \ + \ a_{n-1} \  \partial_{t}^{(n-1)}(
\partial_{tt}-c_{n-1}^2 \partial_ {xx} ) \ + \ ..
 \eeq
\hspace*{2cm}\[... a_{0} \ (
\partial_{tt}-c_{0}^2 \partial_ {xx} )  \]

\vspace{3mm}
\noindent
where  $ a_k \ (k=0..n-1)$ are positive constants.

According to the value of $n$, (\ref{11}) describes several physical
phenomena. As example, when $n=1$, ${\cal L}_n $ can be found
in dynamic of relaxing gases, in magnetohydrodynamics,
 in hereditary electromagnetism (see \cite{r3} and references therein)  and in
isotropic viscoelasticity where, (\ref{11}) models the evolution of
the
 Standard Linear Solid (S.L.S.) (see, f.i. \cite{h,hu}).

In all these models $c_k$'s represent the characterized speeds depending on the
materials properties of the medium and in many physical problems it
results
   $c_0^2 \leq c_{1}^2..\leq c_{n-1}^2 \leq c_n^2$     as it's typical of {\it
wave hierarchies}. \cite{w}.

When $n=1$, the operator (\ref{11}) is stricty-hyperbolic and
 it has been
widely analized in \cite{r3}. It's  fundamental solution ${\cal E}_1$ has been
explicitly determined and singular perturbation problems, together with asymptotic
properties, have been estimated.

Aim of the paper is to draw generalizations of the wide analysis
related to S.L.S. to the case of (\ref{11}) with $n $  arbitrary. For
this,
 a {\em conditioned} equivalence between (\ref{11}) and
  an integro -
 differential operator $\cal M$  related to an appropriate memory
 function $g_n(t),$ is considered.
Owing to  the hypotheses of fading memory, every function $g_n(t)$ can be
approximated by Dirichlet polynomials  \cite{b,s} with appropriate
restrictions on the coifficients of this expansion.
These limitations need that the differential operator  ${\cal L}_n $
 is typical of {\em wave hierarchies} \cite{g,g2}.

By this equivalence, whatever $n $ may be, the fundamental solution
${\cal E}_n $ of (\ref{11}) is explictly achieved and it is estimated in
terms of
${\cal E}_1 $.
 So the maximum properties and asymptotic estimates established
 for the Standard Linear Solid  can be applied to operator
 ${\cal
L}_n $  defined in (\ref{11}).
Moreover, boundary layer problems, typical of dissipative media, can be
rigorously estimated. \cite{dm,dmr}.

These results are applied to  {\em the Rouse model} and
 {\em the reptation model } which describe different aspects of polymer
chains  and have met with reasonable success.\cite{f}-\cite{de}.

\vspace{8mm}

 \section{
{\bf \hspace*{-8mm}. \hspace*{-2mm} DIFFERENTIAL CONSTITUTIVE EQUATION}
}
 \setcounter{equation}{0}

 \hspace{5mm}

Let ${\cal B} $  a linear, isotropic,
 homogeneous system and let  $\underline {u} (\underline {x},t)$   the
 displacement field  from an undeformed reference
 configuration  ${\cal B}_0$. If  $\underline {u} \; = u(x,t) \;
 \underline {i}$ and $\rho_0 $
 denotes the mass density in ${\cal B}_0$,
 the equations of one-dimensional motions of ${\cal B}$ are

\vspace{3mm}
 \beq                                   \label{21}
\rho_0 \; u_{tt} = \sigma  + f , \ \ \   \varepsilon= u,
 \eeq

\vspace {3mm}
\noindent
where $\underline{f}= f  \underline{i}$  is the known body force,  while $\sigma $ and
$\varepsilon $ are the only non vanishing components of the stress and
the strain tensors.

When the viscoelastic behavior of $\cal B$ is of {\it rate-type}, the
well known stress-strain constitutive relation is

\vspace{3mm}
 \beq                                   \label{22}
\sum _{k=0}^n a_k \ \partial_t^{k} \sigma =\sum _{k=0}^n \alpha_k
\ \partial_t^{k} \varepsilon
 \eeq

\vspace{3mm}\noindent
with $a_k, \alpha_k$ constant $(a_n,\alpha_n \neq 0)$.

Then, by (\ref{21}), (\ref{22}), the displacement field  ${\underline
u}(\underline{x},t)$  is solution of the higher order equation like:

\vspace{3mm}
 \beq                                   \label{23}
{\cal L}_n v \equiv \sum _{k=0}^n a_k \ \partial_t^{k} (
v_{tt}-c_{k} v_ {xx} ) =F
 \eeq

\vspace{3mm}\noindent
where:

\vspace{3mm}
 \beq                                   \label{24}
c_k= \alpha _k / \rho_0 a_k, \ \ \ F=(1 / \rho_0 ) \ \sum _{k=0}^n a_k
\ \partial_t^{k} f.
 \eeq

\vspace{3mm}
The constitutive relation (\ref{22}) includes various classical
mechanical models, as  Maxwell and  Kelvin - Voigt models \cite {c}.
Moreover, when $n=1$ and

\vspace{3mm}
 \beq                                   \label{25}
0  <   c_0  <  c_1 , \ \ \ \eta = a_1/a_0 > 0
 \eeq

\vspace{3mm}\noindent
one has the case of the {\it Standard-Linear  Solid}  (S.L.S.)  which is
modelled by the strictly- hyperbolic third order equation:

\vspace{3mm}
 \beq                                   \label{26}
{\cal L}_1 v \equiv \eta \  \partial_t (
v_{tt}-c_{1} v_ {xx} ) + \ v_{tt}-c_{0} \; v_
{xx} = (1/a_0)F.
 \eeq

\vspace{3mm}
The fundamental solution ${\cal E}_1$ of this operator has been obtained  in
\cite{r3}, also when $\underline{{\bf x}} \in \mit{R}^2$ or
$\underline{\bf{x}} \in
\mit{R}^3$.  Further, numerous basic properties of ${\cal E}_1$ have been
rigorously estimated and the wave behavior of S.L.S. is now acquired.

\vspace{3mm}
When $n > 1$ and all the $c_k$' s are positive, then waves of different
orders appear and their roles must be clarified in order to see how
each set is modified by the presence of the other. Obviously,
 wave or dispersive behavior depend on the requirements of the
coefficients $a_k$ and $c_k$ due to physical properties of the system $\cal
B$.

For this, we will analyze the restrictions imposed on the constants
$a_k$ and $c_k$ by usual hypotheses of fading memory for $\cal B $.

\vspace{8mm}

\begin{center}
 \section{
{\bf \hspace*{-8mm}. \hspace*{-2mm}INTEGRAL CONSTITUTIVE EQUATION}
}
\end{center}
 \setcounter{equation}{0}

 \hspace{5mm}

When the strain amplitudes are not too large, the behavior of most
viscoelastic media is fairly well modelled by linear hereditary
equations like:

\vspace{3mm}
 \beq                                   \label{31}
\varepsilon (t) = J(0) [\sigma(t) + \int_{-\infty}^t \dot J (t-\tau)
\sigma (\tau) d\tau] ,
 \eeq

\vspace{3mm}\noindent
where $J(t)$ denotes the creep-compliance and the integral term needs
the knowledge of the past history of the stress.

Usually, according to {\it fading memory} hypotheses, $\dot J(t)$ is a
positive fast decreasing function. For instance, several real
materials as polymers, rubbers, bitumines, have satisfactory
representations by means of chains of S.L.S. elements in series or
parallel \cite{hu,f}. In the series case, the creep function is

\vspace{3mm}
 \beq                                   \label{32}
 J_n(t) = J_n(0) [ 1 + \sum _{k=1}^n \frac{B_k}{\beta_k}(1-
 e^{-\beta_kt})],
 \eeq

\vspace{3mm}\noindent
where $n$ is the number of elements in the chain, $\tau _k= \beta_k^{-1}$
are the characteristic times and $J_n(0)  $ denotes the
elastic compliances.

Then, if one puts:

\vspace{3mm}
 \beq                                   \label{33}
c^2 = [\rho_0 J_n(0)]^{-1} , \ \  F_* =c^2 [J_n(0) f
+\int_{-\infty}^0 \dot J_n (t-\tau)
\sigma_x (\tau) d\tau] ,
 \eeq

\vspace{3mm}\noindent
by (\ref{21}), (\ref{31}), (\ref{32}) one deduces:

\vspace{3mm}
 \beq                                   \label{34}
 {\cal M}u \equiv c^2 u_{xx}- u_{tt} - \int_0^t g(t-\tau)u_{\tau\tau}
 d\tau = - F_*(x,t),
 \eeq

\vspace {3mm}
\noindent
with

\vspace{3mm}
 \beq                                   \label{35}
g \ = \ g_n(t)= \sum _{k=1}^n B_k e ^{-\beta _k t} \ = \dot J_n(t)
/J_n(0).
 \eeq

\vspace{3mm}
\noindent
In this memory function, $n $ is quite
 arbitrary and constants $B_k $  and frequencies $\beta_k$
are such that:

\vspace{3mm}
\beq            \label{36}
0 < \beta _1 < \beta_2
...< \beta_n; \ \ \ \  B_k > 0 \ \ \ \ \ \forall k =1,2....n.
\eeq

\vspace{3mm}

These hypotheses assure that:

\vspace{3mm}
 \beq                                   \label{37}
g(t)>0, \ \ \dot g < 0 , \ \  \ddot g > 0,  \ \ \forall t \geq 0
 \eeq

\vspace{3mm}
\noindent
according to the convexity assumption considered by Dafermos \cite{da}.

We observe that the representation (\ref{35}) of the memory function
 $g$ is not restrictive because
well-known Muntz and Schawart'z theorems
 \cite{b,s} imply that whatever $C^0 ({\mit R }^+)$ function can be
 uniformly represented
 by means of Dirichlet polynomials.
 Moreover, as $n$ is arbitrary, the constants $B_k,\beta_k$ can be determined in order to fit
the experimental curves for $g(t)$ to any prefixed degree of
approximation \cite{f}.

By (\ref{32}), (\ref{36}) one has

\vspace{3mm}
 \beq                                   \label{38}
 J_n(\infty) = J_n(0) [ 1 + \sum _{k=1}^n \ \frac{B_k}{\beta_k}
] > J_n(0).
 \eeq

\vspace{8mm}

 \section{
{\bf \hspace*{-8mm}. \hspace*{-2mm} FADING MEMORY
 AND WAVE HIERARCHIES}
}
 \setcounter{equation}{0}

 \hspace{5mm}

The initial data related to
(\ref{23}) and
(\ref{34}) let be null and let

\vspace{3mm}
\beq                                   \label{41}
P(s) = \sum_{k=0}^n \ \mu _k \ s^k \ \ \  \ , \ Q(s)= \sum_{k=0}^n \lambda_k
 \ s^k
\eeq

\vspace{3mm}\noindent
with

\vspace{3mm}
\beq                                   \label{42}
\mu _k = a_k/a_n \ \ \ \ \ , \ \lambda_k = a_kc_k/a_nc_n \ \ \ \
(k=0,..n)
\eeq

\vspace{3mm}\noindent
so that $\mu_n = \lambda_n=1.$

\noindent
Further, let

\vspace{3mm}
\beq                                   \label{43}
G(s)=  \sum
_{k=1}^n \frac{B_k}{s+\beta_k}
\eeq

\vspace{3mm}
\noindent
the Laplace transform of the memory function (\ref{35}).

Then, if one applies the Laplace transformation to (\ref{23}) and
(\ref{34}), it results:

\vspace{3mm}
\beq                                   \label{44}
\hat{v}_{xx} - \frac{s^2}{c_n} \ \ \frac{P(s)}{Q(s)} \hat{v} =
 \ - \frac{1}{a_nc_n} \ \ \frac{\hat{F}}{Q(s)}
\eeq

\vspace{3mm}
\beq                                   \label{45}
\hat{u}_{xx} - \frac{s^2}{c^2} \ \ [1+G(s)] \hat{u} = \
- \frac{\hat{F}_*}{c^2}
\eeq

\vspace{3mm}\noindent
where
$(\hat{\ \ })$ denote the {\it L} -transform of ( ).

By comparing (\ref{44}), (\ref{45}) one deduces

\vspace{3mm}
\beq                                   \label{46}
\frac{P(s)}{Q(s)}=  \frac{c_n}{c^2} \ \ [1+G(s)]
\eeq

\vspace{3mm}\noindent
and the
 polinomial identity implies
$c_n=c^2$ and

\vspace{3mm}
\beq                    \label{47}
\left\{
\begin {array}{l}
\lambda_0 =
 \beta_1 \beta_2..\beta_n \\
..............................\\
\lambda_{n-2}=
\beta_1 \beta_2+ \beta_1 \beta_3+..\beta_{n-1} \beta_n  \\
\lambda_{n-1}=    \beta_1+....+\beta_n
\end {array}
\right.
\eeq

\vspace{3mm}
\noindent
So, owing to (\ref{36}), all the $\lambda_k$'s are positive. Further, as
for $\mu_k$, one has:

\vspace{3mm}
\beq                    \label{48}
\left\{
\begin {array}{l}
\mu_0= \lambda_0 +   B_1
(\beta_2..\beta_n)+...B_n(\beta_1..\beta_{n-1})\\
..............................\\
\mu_{n-2} = \lambda_{n-2} +  B_1
(\beta_2+..+\beta_n)+...B_n(\beta_1+..+\beta_{n-1})  \\
\mu_{n-1} = \lambda_{n-1} +    B_1+....+B_n
\end {array}
\right.
\eeq

\vspace{3mm}\noindent
 and (\ref{36}), (\ref{48}) imply too:  $ 0<\lambda_k<\mu_k \ \ (k=0,..n-1)$. As
 consequence:

\vspace{3mm}
\beq                        \label{49}
 0< c_k < c_n = c^2 \ \ \ \ \ (k=0,...n-1)
\eeq

\vspace{3mm}
\noindent
At last, by (\ref{47}),(\ref{48}), it follows:
$\frac{\lambda_0}{\mu_0} \, < \,\frac{\lambda_1}{\mu_1} \,<
 \, \frac{\lambda_{n-1}}{\mu{n-1}}$  and so

\vspace{3mm}
\beq                                   \label{410}
0 <  c_0 < c_1 ....\ < c_n.
\eeq

\vspace{3mm}
\noindent
So, the following property holds:

\vspace{3mm}
{\bf Property 4.1.} {\em Hypotheses of fading memory} (\ref{35})
(\ref{36}) {\em imply
that the differential operator} (\ref{23}) {\em is typical of wave
hierarchies}.   \hbox {} \hfill \rule {1.85mm} {2.82mm}

\vspace{3mm}
Vice versa, the inverse transformation of  (\ref{47}),(\ref{48})
requires carefulness.

\noindent
When the differential equation  (\ref{23}) is prefixed, in order to
obtain the dual hereditary equation  (\ref{34}) with a memory function
$ g(t)$ satisfying  (\ref{35}), (\ref{36}), appropriate restrictiones on
the constants $a_k, c_k $ must be imposed.

At first,  (\ref{43}), (\ref{46}) imply

\vspace{3mm}
\beq                                   \label{411}
\frac{P(s)}{Q(s)} \  =  \ B_0  + \sum
_{k=1}^n \frac{B_k}{s+\gamma_k}
\eeq

\vspace{3mm}
\noindent
where all the roots $s=-\gamma_k$  di  $Q(s)$ are real and simple, with
$\gamma_k >0.$ Moreover the conditiones $B_k>0,$ which are sufficient to
verify (\ref{37}), involve further limitations.

{\bf Example 4.1} - When $n=1$, one has:  $c^2=c_1, B_0=1$, and

\vspace{3mm}
\beq                              \label{412}
\beta_1  =\frac{a_0c_0}{a_1c_1}>0  \ \ \ , B_1 =
\frac{a_0}{a_1}(1-\frac{c_0}{c_1}) >0
\eeq

\vspace{3mm}
\noindent
which represent the known restrictions typical of S.L.S.  \hbox {} \hfill \rule {1.85mm} {2.82mm}

\vspace{3mm}
{\bf Example 4.2} - When $n=2$, one has: $c^2=c_2, B_0=1$, and
$\beta_ 1 , \beta_2 $ are real iff:

\vspace{3mm}
\beq                          \label{413}
\omega^2 = (a_1c_1)^2 - 4 (a_0c_0)(a_2c_2)>0
\eeq

\vspace{3mm}
\noindent
Then, it results:

\vspace{3mm}
\beq                          \label{414}
\beta_1 = \frac{1}{2a_2c_2} (a_1c_1 - \omega) , \ \ \ \beta_2 =
\frac{1}{2a_2c_2} (a_1c_1 + \omega)
\eeq

\vspace{3mm}
\noindent
so that $0<\beta_1<\beta_2$ . Further

\vspace{3mm}
\beq                          \label{415}
B_i = \frac{(-1)^{i-1}}{\omega} [a_0 (c_2 - c_0 ) - a_1 \beta_i
(c_2-c_1 )]. \ \ \ \ \\ (i=1,2)
\eeq

\vspace{3mm}
Thus, it is $ B_1>0, B_2>0 $ iff

\vspace{3mm}
\beq                          \label{416}
\beta_1 < \frac{a_o}{a_1} \ \ \frac{c_2-c_0}{c_2-c_1} < \beta_2.
\eeq

\vspace{3mm}
Therefore, the fourth-order operator

\vspace{3mm}
\beq                          \label{417}
a_2 (u_{tt}-c_{2} u_ {xx} )_{tt} + a_1 ( u_{tt}-c_{1} u_{xx})_{t}+a_0
(u_{tt}-c_0 u_{xx})
\eeq

\vspace{3mm}
\noindent
can be analyzed by (\ref{34}), (\ref{35}) , (\ref{36}) when the
constants $a_k, c_k$ satisfy  (\ref{413}) and (\ref{416}).
\hbox {} \hfill \rule {1.85mm} {2.82mm}

\vspace{8mm}

\begin{center}
 \section{
{\bf \hspace*{-8mm}. \hspace*{-2mm} ESTIMATES FOR THE HEREDITARY MODEL}
}
\end{center}
 \setcounter{equation}{0}

 \hspace{5mm}

Let ${\cal B}_n $ the viscoelastic model characterized by the memory
function $g_n $ in (\ref{35}); the case $n=1$ corresponds to the L.S.L.
${\cal B}_1. $

In \cite{r1,r2}, the fundamental solution $E_n$ of the operator
${\cal M} $ in (\ref{34}) has been explicitly determined, whatever $n $ may
be.
 If   $\eta$ is the step -
 function and  $I_0$  is the modified Bessel function of first kind, it
results:

\vspace{3mm}
\beq                \label{51}
E_n=E_n(\beta_1..\beta_n,B_1..B_n)=\frac{1}{2c} \eta(t-r) (A_1+A_2)
\eeq

\vspace{3mm}
\noindent
with

\vspace{3mm}
\beq                \label{52}
A_1 = e^{-g_0 t/2} I_0 (\frac{ g_0}{2} \ \sqrt{t^2-r^2})
\eeq

\vspace{3mm}
\beq                \label{53}
A_2 = \frac{1}{\pi} \int_0^\pi d\theta \int_r^t e^{-g_0 z} H ( z,t-w)
du,
\eeq

\vspace{3mm}
\noindent
and
 $g_0= g(0), r=|x|/c, \ \ 2z =w-cos\theta \ (w^2-r^2)^{1/2}.$
 Futher, if

\vspace{3mm}
\beq                \label{54}
\phi_k(z,t)=e^{-\beta_kt} \sqrt{B_k \beta_k z/t} \ \ I_1 (2\sqrt{B_k
\beta_k z \ t}),
\eeq

\vspace{3mm}
\noindent
one has:

\vspace{3mm}
\beq                 \label{55}
H(z,t)= \sum \ \phi _k + \sum _{k_1,K_2} \phi_{k_1} *\phi_{k_2}+...
\eeq

\vspace{3mm}
\noindent
where sums are computed according to the simple combination of the
indices  $k_1  k_2.. k_n$  and $*$  denotes the convolution with respect to t.

\vspace{3mm}
Moreover the fundamental solution $E_n$ related to
 ${\cal B}_n $ and defined in (\ref{51})-(\ref{55}), can be rigorously
 estimated in terms of the fundamental solution $E_1$ related to an
 appropriate S.L.S.  ${\cal B}_1^* $ defined by

\vspace{3mm}
\beq             \label{56}
g_1 = b \  e^{-\beta_1 t} \ \   \
  with \ \ b = \beta_1  \sum_1^n
\frac{B_k}{\beta_k}.
\eeq

\vspace{3mm}
In fact, if
 $\Gamma $ is the open forward
characteristic cone $\{ (t,x) : t >0 \ |x| < ct \}$, and  $\chi_n =
 \prod_{k=2}^n \ (\frac{B_k}{\beta_1})^2$,
then the following theorem holds:

\vspace{3mm}
{\bf Theorem 5.1} - {\em If the memory function is given by } (\ref{35})
(\ref{36}), {\em
then the fundamental solution} $E_n$ {\em of} $ {\cal M}$  {\em is a never negative }
$C^{\infty}(\Gamma)$ {\em function and
 it
satisfies the estimate}:

\vspace{3mm}
\beq                     \label {57}
0< E_n(\beta_1..\beta_n,B_1..B_n)< \chi_n \
\ E_1(\beta_1, b
),
\eeq

\vspace{3mm}
\noindent
{\em everywhere in } $\Gamma$
{\em and
whatever } $n$ {\em may be}. \hbox {} \hfill \rule {1.85mm} {2.82mm}

\vspace{3mm}
{\bf Remark 5.1} - The model  ${\cal B}_1^*  $ defined by (\ref{56}) is
physically meaningful.

\noindent
In fact the memory function $g_1$ is related just to the obliviator
because
 $\tau_1= \beta_1^{-1}$ is the longest characteristic time.

Furthermore  ${\cal B}_n$   and   ${\cal B}_1^*$ verify the same
hypotheses of fading memory and  by (\ref{35})
 (\ref{56}), it results:

\vspace{3mm}
\beq             \label{58}
\int _0 ^ \infty g_n(t) dt \ \ = \int _0 ^ \infty g_1(t) dt \ \ = \sum_1^n
\frac{B_k}{\beta_k}.
\eeq

\vspace{3mm}
\noindent
 Moreover, the integral (\ref{58}) affects the
asymptotic analysis of hereditary equation \cite{g,lf}. \hbox {} \hfill \rule {1.85mm} {2.82mm}

\vspace{3mm}
{\bf Remark 5.2.} - By known properties of asymptotic behaviour of
convolutions, the constitutive relation
(\ref{31}) implies:

\vspace{3mm}
\beq                \label{59}
\lim_{t \rightarrow \infty} \varepsilon (x,t)=
J_n(0) \ [1+\int_0 ^\infty g_n (t) dt ] \ \lim_ {t \rightarrow \infty} \sigma
(x,t)
\eeq

\vspace{3mm}
\noindent
provided that $\lim_ {t \rightarrow \infty} \sigma$ exists. Then   ${\cal
B}_n$   and the model   ${\cal B}_1^*$ exhibit the same  asymptotic
behaviour (\ref{59})

Further, by (\ref{35}),  (\ref{38}) it results:

\vspace{3mm}
\beq                \label{510}
\frac{J_n(\infty)}{J_n(0)} =1+ \sum_1^n
\frac{B_k}{\beta_k} = \frac{J_1(\infty)}{J_1(0)}
\eeq

\vspace{3mm}
\noindent
and so, (\ref{510}) implies  $J_n(\infty) \sigma(\infty) = \varepsilon(\infty).$
Consequently, when $t$ is large,
 the behaviour  of  ${\cal B}_n $ is typical of an elastic material
with modulus $J_n(\infty)$. \hbox {} \hfill \rule {1.85mm} {2.82mm}

\vspace{8mm}

 \section{
{\bf \hspace*{-8mm}. \hspace*{-2mm} ESTIMATES RELATED TO WAVE
HIERARCHIES}
}
 \setcounter{equation}{0}

 \hspace{5mm}

When the operator  ${\cal L}_n    $  is reduced to the hereditary
operator  ${\cal M}     $  of (\ref{34}), then estimates of
 Theorem 5.1  can be applied to wave
hierarchies.
Obviously, the equivalence is conditioned by inverse transformation of
(\ref{47})- (\ref{48}) together with  (\ref{36})
(see n.4).

Let  ${\cal L}^*_1    $ the operator (\ref{26}) related to the S.L.S.
  ${\cal B}_1^* $ characterized by

\vspace{3mm}
\beq                     \label{61}
 \eta =\frac{1}{\beta_1+b}, \ \ \ a_0= \frac{\beta_1+b}{\beta_1},
\ \ \ c_0 =\frac{c^2\beta_1}{\beta_1 +b} \ \ \ c_1=c^2.
\eeq

\vspace{3mm}
Now, let  ${\cal L}_n    $ the differential operator given by (\ref{23})
whatever $n $ may be, and let  ${\cal P}_n    $ a prefixed boundary value
problem  related to  ${\cal L}_n    $. The meaningful aspects of
qualitative analysis of the solution of  ${\cal P}_n    $  can be
obtained by means of Theorem 5.1 and by the known properties
  of  ${\cal L}^*_1    $
 \cite{r3}.

So  maximum properties,
asymptotic behaviour, boundary layer estimates, etc. for the solution
of  ${\cal P}_n    $ are deduced by analogous properties related to
${\cal L}^*_1.    $

Moreover, owing to the equivalence between   ${\cal L}_n    $ and  ${\cal
M}     $, it is possible to have explictly the fundamental solution of
  ${\cal L}_n    $ for all $n$. In fact it suffices to apply the
  explicit formula (\ref{51})-(\ref{53}).

As an example the case of {\em polymeric
materials} can
be considered.

{\bf Example 6.1} -
Polymeric materials are very flexible and are easily
formed into fibres, thin films, additives for oils, etc. So theirs
applications to concrete problems are numerous.\cite{d,m,sm}. According to
theories of linear viscoelasticity, two models, that descibe
 different
aspects of polymer chains, have met a reasonable
success:
 {\em the Rouse model} and {\em the reptation
model} \cite{de}.

In both cases the memory function $g(t) $ assumes a form like
(\ref{32})(\ref{33}).
In fact in {\em the reptation model}, the stress relaxation
function is:

\vspace{3mm}
\beq             \label{62}
g(t) = k \sum _{h=0}^n  \frac{1}{(2h+1)^2} \ \ e^{-(2h+1)^2
\frac{t}{\tau_d}},
\eeq

\vspace{3mm}
\noindent
where $k$ is a constant depending
on the polymer physics and the value of the "reptation" time  $\tau_d$ can be fixed according
to
 elasticity experiments \cite {f}.

When the  viscoelastic behaviour is represented by
 {\em the Rouse model}, memory function $g(t)$ is given by
:

 \vspace{3mm}
\beq             \label{63}
g(t) = k_1 \sum _{h=1}^n  e^{2h^2 \frac{t}{\tau_1}},
\eeq

\vspace{3mm}
\noindent
where
 the relaxation time $\tau_1$  can be calculated by means of
 experimental
results.\cite{de}.

So, if one considers the first two steps in  {\em
the reptation model}, it results:  $B_1=k, \ \ B_2 =B_1/9, \ \ \beta_1=1/\tau_d, \ \
\beta_2= 9\beta_1.$  Consequently the operator (\ref{417}) is
characterized by constants:

\vspace{3mm}
\beq               \label{64}
\left\{
\begin {array}{lll}
c_0= c^2 \
\frac{81}{81+82k \tau_d} &
 c_1= c^2 \ \frac{9}{9+k\tau_d}
& c_2= c^2  \\
                            \\
a_0=1+ \frac{82}{81} \, k \tau_d &
a_1=\frac{10 \tau_d^2}{9}(\frac{1}{\tau_d}+\frac{k}{9}) &
a_2 =\frac{\tau_d^2}{9}.
\end {array}
\right.
\eeq

\vspace{3mm}
\noindent
Analogously, in {\em the Rouse model}, beeing $B_1=B_2=k_1,
\ \ \beta_1=2/\tau_1, \ \  \beta_2=4\beta_1$, one has:

\vspace{3mm}
\beq               \label{65}
\left\{
\begin {array}{lll}
 c_0= c^2
\frac{8}{8+5 k_1 \tau_1}  & c_1= c^2  \frac{5}{5+ k_1 \tau_1}
& c_2= c^2 \\
                             \\
 a_0 = 1+ \frac{5 k_1 }{8}\tau_1  &
a_1 = \frac{\tau_1^2}{16} (2 k_1 +\frac{10}{\tau_1}) &
a_2=\frac{\tau_1^2}{16}.
\end {array}
\right.
\eeq

\vspace{3mm}
The {\em wave hierarchies} defined by
 (\ref{64}) or (\ref{65})  are governed by the operator   ${\cal L}^*_1    $
of the Standard Linear Solid defined, respectively, by:

\vspace{3mm}
\beq               \label{66}
\left\{
\begin {array}{llll}
c_0= c^2 \
\frac{81}{81+82k \tau_d} &
 c_1= c^2  &
a_0=1+ \frac{82}{81} \, k \tau_d &
\eta =\frac{81 \tau_d}{81+82 k \tau_d}, \\
 \\
 c_0= c^2
\frac{8}{8+5 k_1 \tau_1}  &
 c_1= c^2 &
 a_0 = 1+ \frac{5}{8} k_1 \tau_1 &
\eta = \frac{4 \tau_1}{8+5 k_1 \tau_1}.
\end {array}
\right.
\eeq

 \hbox {} \hfill \rule {1.85mm} {2.82mm}

These results have been confirmed also in \cite{im} for entangled
polymers with chain strecth.

\vspace{10mm}
\noindent
\vspace{5mm}
\begin{center}
\bf{REFERENCES}
\end {center}

\end {document}